# Influence of synoptic and local atmospheric patterns on PM10 air pollution levels: a model application to Naples (Italy)

Alberto Fortelli[1], Nicola Scafetta[2], Adriano Mazzarella[2+]

[1] Meteorological Network Campanialive, Naples, Italy.

[2] Meteorological Observatory, Department of Earth Sciences, Environment and Georesources, University of Naples Federico II, Largo S. Marcellino, 10 - 80138 Naples, Italy.

[+]Corresponding Author: adriano.mazzarella@unina.it

ABSTRACT

We investigate the relationship between synoptic/local meteorological patterns and PM10 air pollution levels in the metropolitan area of Naples, Italy. We found that severe air pollution crises occurred when the 850 and 500 hpa geopotential heights and their relative temperatures present maximum values above the city. The most relevant synoptic parameter was the 850 hPa geopotential height, which is located about 1500 m of altitude. We compared local meteorological conditions (specifically wind stress, rain amount and thermal inversion) against the urban air pollution levels from 2009 to 2013. We found several empirical criteria for forecasting high daily PM10 air pollution levels in Naples. Pollution crises occurred when (a) the wind stress was between 1 and 2 m/s, (b) the thermal inversion between two strategic locations was at least 3°C/200m and (c) it did not significantly rain for at least 7 days. Beside these meteorological conditions, severe pollution crises occurred also during festivals when fireworks and bonfires are lighted, and during anomalous breeze conditions and severe fire accidents. Finally, we propose a basic model to predict PM10 concentration levels from local meteorological conditions that can be easily forecast a few days in advance. The synthetic PM10 record predicted by the model was found to correlate with the PM10 observations with a correlation coefficient close to 0.80 with a confidence greater than 99%. The proposed model is expected to provide reliable information to city officials to carry out practical strategies to mitigate air pollution effects. Although the proposed model equation is calibrated on the topographical and meteorological conditions of Naples, it should be easily adaptable to alternative locations.

# 1. Introduction

Air pollution is identified by the World Health Organization (WHO) as responsible for several million deaths per year (Abbey et al., 2005; Beelen et al., 2008; Crosignani et al., 2002; Dockery et al., 1993; Laden et al., 2006; Pruss-Ustun et al., 2016). In most industrialized countries, including Italy, particulate matter emissions are regulated by severe environmental laws because smog crisis can involve large territories and affect their residents. Several studies carried out in the U.S.A. and in Europe points out high risks for lung cancer due to thin particulate matter (e.g. Vineis et al., 2006).

This work aims to investigate the meteorological conditions that can favor a smog crisis in a densely populated urban area and to provide a semi-empirical model able to predict the PM10 concentration starting from meteorological parameters such as wind, rain and vertical layer instability. The possibility of forecasting the conditions yielding a probable particulate crisis could be efficiently used by city policymakers to implement regulations to mitigate the phenomenon.

In this study we investigated the metropolitan area of Naples, which is located in Central Italy on the Tyrrhenian coast. The city of Naples is affected by an increasing urban heat island (Palumbo and Mazzarella, 1981; Mazzarella and Giuliacci, 2011) and is characterized by recurrent periods of high pollution levels due to the particular morphology of the region and to relevant anthropogenic emission sources (Adiletta et al., 1981; Palumbo and Mazzarella, 1981).

The air dust pollutant that we investigated is the particulate matter with a diameter of 10μ or less, known as PM10. This kind of particulates can be produced by natural phenomena such as dust storms, forest and grassland fires and sea spray. However, the major contribution to PM10 concentration in a large city derives from human activities such as the burning of fossil fuels in vehicles, in airplanes, in power plants and in other industrial processes. In the case of a major seaport city such as Naples, a significant contribution is also due to the large cruises and commercial ships that regularly visit the city (cf. Mueller et al., 2011). These large ships operate continuously also when anchored in the port emitting large amount of pollutant. An additional component of PM10 pollution derives from the continuous erosion of the urban environment (streets and buildings) due to weather conditions and anthropogenic traffic.

Italian regulations require that the number of days per year in which PM10 could exceed 50 µg/m$^3$ have to be less than 35 (Gazzetta Ufficiale, 2010). To reduce such occurrences, the traffic circulation and all city sources of polluting emissions must be limited.

Weather factors govern the dynamics and thermodynamics of the atmosphere and of the air pollutant concentration (Aldrin and Hoek, 2005). The main meteorological parameters are wind stress and direction, vertical thermal gradients, atmospheric humidity, rain and solar radiation. For example, wind absence and air thermal inversion block the convective motions of the air, thus inhibiting vertical mixing and causing an accumulation of pollutants in the lower atmosphere layers (Kallos et al., 1991; Kelessis et al., 2009).

Urban plume is usually forecast according to generic deterministic complex dispersion models involving multi-parameters related to air pollution, traffic volume, meteorology and local topography (Aldrin and Haff, 2005; Binkowski and Roselle, 2003; Finardi et al., 2008; Silibello et al., 2008; Wang et al., 2004; Wang et al., 2008; Xu, et al., 2008; Wang et al., 2010; Zhang et al., 2004). The characteristics of these analytic models, however, make them difficult to adapt to sites with a complex morphology. The goal of this work is to provide a simpler basic and versatile model made of the least number of degrees of freedom able to predict with a good approximation the PM10 concentration. Such a model also makes the contribution of the various meteorological parameters more easily intelligible. Moreover, because the forecast of meteorological parameters is promptly available from National Weather Forecasting Centers (e.g. GFS, ECMWF, UKMO, etc.), the proposed model could be used to easily predict a PM10 crisis in a large city.

## 2. Physics of the process

Ground level air pollution depends on the particulate emission rate and on the following weather parameters.

2.1 Wind stress

Pollutants are mechanically dispersed by wind stress. Thus, wind stress should be the most important meteorological factor in determining PM10 air concentration in a city (Holton, 2004).

2.2 Vertical air temperature profile

Vertical distribution of air temperature determines the atmospheric stability controlling the depth of vertical mixing (Leighton and Spark, 1997; Holton, 2004; Li et al., 2015). The boundary layer of the vertical mixing region governs the air volume that confines near ground-level emitted pollutants. The best conditions for air pollution dispersion occur with strong air instability and high mixing while strong negative thermal gradient favors the air mixing and the diffusion of the pollutants (Holton, 2004).

2.3 High pressure and rain amount

High pressure is usually characterized by fair weather, that consists in light winds, lack of storms and of precipitations, and nearly cloud-free sky. Such a condition can be of particular concern for air quality because of the nighttime inhibition of air motion both vertically and horizontally (Holton, 2004). High pressure usually implies lack of rain and this favor the deposit of particulate which may be lifted up by the wind and become an additional factor to city emissions.

In the following we propose a semi-empirical model linking the major meteorological parameters with air pollution concentration.

## 3. A model for daily PM10 air concentration

Herein we propose a model describing the air pollution ground level concentration $\delta(t)$. This function depends on a number of parameters such as: the average emission per surface unit $\sigma(t)$; the average wind speed $v(t)$; the average air pollution ground level concentration external to the city region $\delta_{ext}(t)$ that is brought on the city by the wind; an estimate of the average vertical height $h(t)$ of the polluted air layer; the average city horizontal dimension d; the average amount of rain per surface unit ρ(t) and the average rain drop size r.

3.1 Basic model

Assuming that the wind and the rain clean the air and that the city air is constrained in a box of sides b, b and h, it is easy to determine that the differential equation of δ(t) is given by:

$$\frac{d\delta(t)}{dt} = \frac{\sigma(t)}{h(t)} + \frac{v(t)}{d}\delta_{ext}(t) - \left(\frac{v(t)}{d} + \frac{\rho(t)}{r}\right)\delta(t) \tag{1}$$

Eq. 1 can be easily solved assuming σ(t), v(t), h(t) and ρ(t) constant within the integration time interval, e.g. 1 day. This gives:

$$\delta(t) = \left(\frac{\sigma}{h} + \frac{v}{d}\delta_{ext}\right) / \left(\frac{v}{d} + \frac{\rho}{r}\right) + \delta(0)e^{-\left(\frac{v}{d} + \frac{\rho}{r}\right)t} \tag{2}$$

,

where δ(0) is the initial value of the PM10 concentration. Because our data are daily means, we average Eq. 1 over one day (86000 s) to get a daily equilibrium PM10 concentration value and obtain the following equation:

$$\delta_{eq} = \left(\frac{\sigma}{h} + \frac{v}{d}\delta_{ext}\right) / \left(\frac{v}{d} + \frac{\rho}{r}\right) + \frac{\delta(0)}{86000}\left(1 - e^{-\left(\frac{v}{d} + \frac{\rho}{r}\right)86000}\right) / \left(\frac{v}{d} + \frac{\rho}{r}\right). \tag{3}$$

Eq. 3 indicates that $\delta_{eq}$ is inversely related to the wind speed and to the high h of the polluted boundary layer, and it is directly related to the horizontal dimension of the city b and to the rate of pollution emission per surface unit. Rain acts similarly to the wind by cleaning the air columns through which rain drops fall. $\delta_{eq}$ also depends on the pollution level of the air $\delta_{est}$ that the wind brings inside the city from the periphery. The latter parameter depends on the industrial/urbanization of the area surrounding the city, on the wind direction (e.g. when the wind comes from the sea, $\delta_{est}$ is lower than when it comes from other directions) and on the amount of possible other dust sources (e.g. from desert dust) reaching the city. $\delta_{eq}$ of a given day also depends on the initial value δ(0) that can be approximated to the $\delta_{eq}$ of the previous day.

Eq. 3 is an useful first approximation model of the process herein discussed. Let us now derive the equations for the vertical depth h and for the emission parameter σ.

3.2 Determination of h

We assume that h depends on the thermal vertical inversion index Δθ (Holton, 2004):

$$h = h_0 e^{-\frac{\Delta\theta}{c_h}} \qquad (4)$$

Large Δθ means severe thermal inversion and in such a situation the pollutant is constrained closer to the surface. Ideally, for Δθ diverging to infinity, h should converge to zero. The parameters $h_0$ and $c_h$ are specific constants related to the characteristics of the location.

We estimate Δθ in the following way. We use the local daily minimum temperature records from the observatories of San Marcellino ($T_M$) (located in the city center at 50 m a.s.l.) and of Sant'Elmo ($T_E$) (located on a hill at almost at 1 km of distance at 250 m a.s.l.) (see Figure 1). Then, we calculate the potential temperatures of Sant'Elmo ($\theta_E$) and of San Marcellino ($\theta_M$) (Holton, 2004) using the following Poisson equation:

$$\theta = T\left(\frac{P_0}{P}\right)^{0.2854}, \qquad (5)$$

where $T$ is $T_M$ or $T_E$ (K) in the investigated stations, respectively, P is the atmospheric pressure (hPa) in the same stations and $P_0$ (hPa) is the reference pressure, which is that measured in San Marcellino. Finally, the inversion index is defined as:

$$\Delta\theta = \theta_E - \theta_M, \qquad (6)$$

Note that because the altitudes of the two observatories differ by about 200 m, Δθ is about $T_E - T_M + 2$. There is thermal inversion when Δθ > 0.

3.3 Determination of σ

The index σ depends on a number of anthropogenic and meteorological factors. Assuming the anthropogenic pollutant emission constant, σ must: 1) increase with the wind stress (v) because wind lifts dust and causes erosion; 2) increase with the number of drought days (Δγ) that causes a continuous accumulation of deposited and suspended dust; 3) decrease to approximately zero as the wind converges to zero because such a condition favors particulate deposition. Suspended dust includes that trapped in the so-called "urban canyons", that is, in the numerous alleyways that characterize the city of Naples which is made mostly of 15-25 m tall buildings quite close to each other.

According to the above expectations, we model σ as:

$$\sigma = \sigma_0 (1 + c_p \Delta\gamma) \frac{1 - e^{-v/c_v}}{1 - e^{-1/c_v}} . \qquad (7)$$

The variable Δγ indicates the number of days without rain necessary to wash off the deposited particulate. Δγ is reset to zero when it rains more than a given amount $\rho_0$. If it rains less than this, the deposited particulate is assumed to be washed off in linear proportion. The variable Δγ is then calculated as:

$$\Delta\gamma_t = \frac{1}{2\rho_0} (\rho_0 - \rho + |\rho_0 - \rho|)(1 + \Delta\gamma_{t-1}), \qquad (8)$$

where $\rho_0$ is a given daily amount threshold of rain (we found $\rho_0$ = 5 mm/day) above which the accumulated dust is completely washed off.

The parameter $c_p$ indicates the increase in particulate emission induced per each day of drought (we found $c_p$ = 2%). The last factor of Eq. 7 is equal to 1 when the wind speed is v = 1 m/s and goes to zero per vanishing v. If v diverges, σ converges to a constant that depends on the parameter $c_v$. This limit indicates the maximum particulate emission amplification due to wind erosion. Finally, the parameter $\sigma_0$ indicates the average particulate emission per unit of time and surface when Δγ = 0 and v = 1 m/s. This parameter can vary with the anthropogenic emissions and can be potentially regulated by policymaker decisions.

Eqs. 4-8 complete Eq. 3, which represents our proposed model for PM10 air concentration inside the city of Naples.

## 4. Meteorological data

We analyzed data referring to 4-month winter intervals (from November to February) of PM10 records from 2009 to 2013 regarding the urban area of Naples provided by daily bulletins (www.arpacampania.it). These PM10 data are averages measured at 8 stations distributed around the city as indicated in figure 1.

The meteorological data herein considered are daily air temperature, atmospheric pressure, rain amount and wind stress measured at the first class Observatory of San Marcellino in the center of Naples (50 m a.s.l., Lat 40°50'48" N Long 14°15'31" E) and at the Sant'Elmo meteorological station (1 km West from San Marcellino and 250 m a.s.l., http://www.campanialive.it/). See Figure 1 for the locations of these two observatories.

The PM10 records provided by ARPAC refer mostly to pollutant concentrations measured at 2-3 meters above the ground street level. On the contrary, the wind stress is measured on the turret of the Observatory of San Marcellino at about 30 m above the ground level. Thus, the measured wind stress is indicative of the air motion above the city while the PM10 concentration values depend also on the air condition within the urban canyons.

We also use the 500 and 850 hPa geopotential height and relative temperature records above Naples provided by the National Center for Environmental Prediction (NCEP) database (www.ncep.noaa.gov).

## 5. PM10 concentration level dependency on synoptic meteorological parameters

First we use an approach involving the dependency of air pollution in Naples on synoptic meteorological parameters (Cheng and Chenh, 2008; Flokas et al., 2009). For this, we compare the PM10 record against the geopotential height of 850 hPa and 500 hPa surfaces and the temperature at same levels provided by NCEP database (Leighton and Spark, 1997; Li et al., 2016; Triantafyllou, 2001). To optimize the comparison we took the four geopotential height records corresponding to the four closest geographical coordinates around Naples (Lat 40°50' 48" N; Long 14°15'31" E; 50 m asl), that is: 40.0N-12.5E, 40.0N-15E, 42.5N-12.5E, 42.5N-15E. Then, using an interpolation, we calculated the geopotential height value exactly above Naples and their relative gradients.

The height of the isobaric surfaces (geopotential) is representative for vertical thermal gradients: high geopotential values correspond to a light vertical temperature gradient, which implies weak vertical air mixing. The gradient of the isobaric surfaces is indicative of wind intensity and direction. Both meteorological conditions are supposed to influence the PM10 concentrations.

Figure 2 depicts the comparison between the geopotential heights at 850 hPa and at 500 hPa and their temperatures against the PM10 records for each of the four analyzed winters. Figures 3 does the same using the gradients. Both sets of figures report the correlation coefficients r between the depicted pairs.

Figure 2 suggests that the PM10 concentration is better correlated with the geopotential heights than with their temperatures. The best correlations ($0.34 < r < 0.65$) are observed using the 850 hPa geopotential height.

Figure 3 suggests that the PM10 concentrations are negatively correlated with the geopotential height gradients better than with those relative to their respective temperatures. The best correlations ($0.45 < |r| < 0.56$) are observed using the gradient of the 850 hPa surface.

Figures 2 and 3 suggest that the PM10 concentration levels on the city of Naples are related to the local vertical thermal distribution and to the local wind stress. In fact, the 850 hPa surface is located at about 1500 m above the sea level and it is just above the surface layer inside which pollution develops. In the following section we relate PM10 concentration levels to local meteorological parameters.

## 6. PM10 concentration level dependency on local scale meteorological parameters

Figure 4 shows the average PM10 record against the wind speed v, the reciprocal of the wind speed 1/v, the thermal gradient $\Delta\theta$, the total daily rain amount $\rho$ and the drought index $\Delta\gamma$ using $\rho_o = 5$ mm/day. The 20 panels also report the cross correlation coefficient r in each case.

During the analyzed 4 periods, the PM10 concentration exceeded the limit of 50 $\mu g/m^3$ for 158 days on 481 days (32.8% of total), while the limit of 100 $\mu g/m^3$ was exceeded for 18 days (3.7% on total), 10 of them clustered in the 2 most relevant smog crises, occurred in November 2009, 18th-21st, and in February 2011, 6th-11th.

The highest correlation values are observed between PM10 and the reciprocal of the wind speed: $0.52 < r < 0.80$. The second highest correlation values are observed between PM10 and the vertical thermal gradient: $0.45 < r < 0.63$. The variable $\Delta\gamma$, indicating the number of drought days, also shows a good correlation with PM10: $0.36 < r < 0.55$. We observe that very high PM10 values always occur during long periods of drought.

Figure 5 depicts in a log-log scale the PM10 index against the wind speed v: the two most correlated parameters. The figure also depicts the data fit (blue) with a function of the type

$$f(v) = a\left(1 - e^{-v}\right)/v \qquad (9)$$

which is a simplification of Eq. 3 where we found $c_v = 1$. Figure 5 suggests that for $v > 1$ m/s, the PM10 concentration decays with the speed, as expected. For $v < 1$ m/s, the PM10 concentration diverges from the $1/v$ behavior and converges to an average maximum of about 100 µg/m³. Moreover, the probability that the PM10 concentration exceeds the 50 µg/m³ limit is about 66% when $v < 1$ m/s.

The 98% of cases where the PM10 concentration is below 30 µg/m³ is observed for $v > 1$ m/s. The 83% of cases where the PM10 concentration is above 100 µg/m³, is observed for 1 m/s $< v <$ 2 m/s. The latter result is reasonable because there should be at least a light wind for placing again in circulation the dust accumulated in the city during the drought days.

## 7. The semi-empirical PM10 model results

Figure 6 shows daily PM10 concentration values against Eq. 3 implemented with Eqs. 4-8. The free parameters of the model are obtained via a simple Monte Carlo simulation to optimize the fit. We found the best results using: $\sigma_0/h_0 = 0.005$ µg/sm³, $c_h = 5$ °C, $c_p = 2\%$ and $c_v = 1$ m/s. In the shown simulations, we assume a constant PM10 emission rate, $\sigma_0$, and that the air outside the city area is clean from all directions, $\delta_{est} = 0$ µg/m³.

The upper panels of Figure 6 use for $\delta(0)$ the $\delta_{eq}$ value calculated by the model for the previous day (model #1), while the lower panels of Figure 6 use for $\delta(0)$ the PM10 value measured the previous day (model #2). Each panel of Figure 6 also reports the correlation coefficients r between the measured and the modeled daily PM10 concentration levels.

For the upper panels, model #1, we found: r = 0.78 for the first period, r = 0.63 for the second period, r = 0.84 for the third period and r = 0.79 per the fourth period. For the lower panels, model #2, we found slightly better correlations: r = 0.79 for the first period, r = 0.67 for the second period, r = 0.85 for the third period and r = 0.80 per the fourth period. Thus, using in Eq. 3 the PM10 values measured on the previous day, the forecast improves slightly implying that this operation would be unnecessary because the model appears sufficiently robust to predict accurate PM10 values.

There are four cases among the most severe PM10 peaks that are not well quantitatively reconstructed by the model. These crises occur on 17-18$^{th}$ January 2010, 1$^{st}$ January 2011, 17-18$^{th}$ January 2011 and 6-11$^{th}$ February 2011. In these cases the PM10 concentration exceeded 130 µg/m$^3$, while the model predicted PM10 concentration values between 50 and 100 µg/m$^3$.

These divergences could be explained in the following way. On January 1$^{st}$ in Naples high PM10 concentration values are expected because the anomalous PM10 emissions produced by fireworks. This occurrence influences the PM10 daily values when the dispersion of the pollutants are limited by meteorological factors.

PM10 peaks with values larger than 100 µg/m$^3$ were observed on January 1$^{st}$, 2011 and 2013 but not on January 1$^{st}$, 2010 and 2012. On January 1$^{st}$, 2010, PM10 level was low (40 µg/m$^3$) because there were in action a) a significant wind stress (v = 4.4 m/s), b) some rain (ρ = 10.4 mm) and c) absence of thermal inversion (Δθ = −0.6 °C), which leaded to a rapid dispersion of the firework pollutants. On January 1$^{st}$, 2011, PM10 level was 133 µg/m$^3$ because of light wind stress (v = 2.0 m/s), no rain for the previous 6 days and a light thermal inversion (Δθ = 0.5 °C). Moreover, there was a significant fire in Naples (https://www.youtube.com/watch?v=gdcuVIYiY4g). On January 1$^{st}$, 2012, the measured PM10 level was moderately high (88 µg/m$^3$) because of a balance between the anomalous firework dust emission and the meteorological physical conditions causing a moderate dispersion (v = 1.5 m/s, no rain for 2 days and moderate thermal inversion Δθ = 1.6 °C). On January 1$^{st}$, 2013 the model well predicted the measured high PM10 level (169 µg/m$^3$, v = 0.5 m/s, no rain for 11 days and significant thermal inversion Δθ = 3.5 °C).

On January 17-18th of every year high PM10 levels should also be expected due to the Saint Antony Abbot festival. During this festival bonfires are lighted all around the city. In 2010 and in 2011 we had 139 and 158 μg/m$^3$ while the model predicted about 55 and 100 μg/m$^3$ respectively because the additional bonfire emission was not modeled.

On February 6-11th, 2011 the largest PM10 crisis of the analyzed period was observed, reaching a value of about 200 μg/m$^3$. In this case, the model predicted high PM10 values but this estimate was about 100 μg/m$^3$. Analysis of the hourly wind stress and direction of those days suggests that the anomalous pollution levels were due to a too moderate and brief sea breeze, which was unable to sufficiently clean the air mass overlaying Naples. In fact, during those days the wind mostly blown from the North-East surrounding areas of Naples that, being almost as densely populated as Naples, produce polluted air. Moreover, on February 9th, 2011, close to the central train station of Naples, a vehicle burned and exploded releasing a large amount of dark smoke (https://www.youtube.com/watch?v=MOCeJMhXq0Q).

Beside the few cases above discussed, in which the model failed to sufficiently predict the observed PM10 concentration values due to the unusual pollutant emission levels of those days, Figure 6 clearly suggests that the proposed model well predicts the PM10 observation with a correlation value close to r = 0.80.

## 8. Discussion and conclusion

Smog crises in urban areas are caused by slight winds that do not facilitate the dispersion of pollutants. In a coastal city like Naples, long periods of low wind stress are more probable in winter times when the strength of thermal circulation (such as breezes) is weaker. During the November to February periods from 2009 to 2013, the city of Naples experiences eight smog crises with PM10 concentrations larger than 100 μg/m$^3$ and a large number of minor crises with PM10 concentrations larger than 50μg/m$^3$.

Meteorological maps representing geopotential height and temperature patterns at 500 hPa and 850 hPa show good correlation with PM10 records. Therefore, a clear relation between the forming and the movement of anti-cyclonic structures and air pollution in Naples appears.

The increase of surface air pressure and the increase in thickness of atmospheric layers between two specific isobaric surfaces, in particular the ones closer to the ground, causes air immobility. The pollutants from industrial and human activities lay stationary above the city (Bahram M. et al., 2014). High pressure system evolution can be forecast several days in advance. This makes possible to infer the conditions favoring the emergence of a possible air pollution crisis several days in advance so that effective mitigation steps can be taken.

However, synoptic meteorological parameters, such as the adopted geopotential values, are correlated to the actual PM10 city concentration values only moderately: the correlation coefficient was best found close to r = 0.50 only for the geopotential height of 850 hPa. The result suggests that local weather parameters need to be taken into account for an improved forecast.

We have compared the measured PM10 city concentration values against local wind stress, rain patterns and a thermal inversion index measured using records collected at the meteorological observatories of San Marcellino and Sant'Elmo, located near the historical downtown of the city.

We found that the PM10 concentration is best correlated with the reciprocal of the wind stress: the correlation coefficient was best found close to r = 0.70. We found that the largest pollution crises (PM10 > 100 µg/m$^3$) usually occurred when the average daily wind speed was between 1 and 2 m/s. For large wind speed (v > 4 m/s), the pollution levels were almost always below 50 µg/m$^3$. For very weak wind (v < 1 m/s), the air pollution rarely exceeded the 100 µg/m$^3$ level suggesting that a significant amount of air pollution derives from erosion and the lifting of deposited dust by the wind.

We also found that PM10 concentrations are negatively correlated with the daily rain amount ( r = -0.30). However, when we compared the pollution level against a drought index, the correlation was more significant, up to r = 0.50. In particular, the worst pollution crises always occurred in concordance of extended drought periods of more than 7 days. The result suggests again that a significant amount of air pollution derives from erosion and the lifting of deposited dust which increase during drought periods and are drastically reduced after sufficiently strong rain (e.g. more than 5 mm/day).

Finally, we found that the volatility of the PM10 record, that is the emergence of its peaks, is significantly correlated with an index Δθ estimating the thermal inversion above Naples (r = 0.60). This thermal inversion index was calculated by comparing the minimum daily temperatures between Sant'Elmo and San Marcellino meteorological stations located at 250 m and 50 m a.s.l., respectively, almost on the same vertical line. Thermal inversion is fundamental in determining the surface pollution air concentration because it determined the width of the vertical air layer where pollution emitted at the surface can diffuse. For thinner layer higher PM10 concentration at the surface is observed. Table 1 reports a qualitative relation between the proposed air temperature inversion index $\Delta\theta$ and the air pollution severity in Naples.

Finally, we proposed a semi-empirical model to calculate the daily PM10 air concentration using wind stress, rain memory and intensity, and thermal inversion (Eqs. 1-8).

The model was able to reproduce the observed PM10 record with a correlation coefficient close to r = 0.80. The only major pollution peaks that were not well reproduced by the model, which in the shown simulations assumes a constant PM10 emission rate, occurred when the pollution emission rate was anomalously high due to fireworks and other bonfires lightened during specific festivals (the January 1st and the January 17-18th Saint' Antony festivals). During another single crisis, on February 6-11th, 2011, the model was able to predict only half of the observed pollution value due probably to an anomalous breeze condition and to a major fire accident occurred in the center of the city.

In general, the proposed model performed quite well in reconstructing the PM10 concentration record and in predicting its peaks. If the model could be implemented with a more realistic time dependent PM10 emission rate $\sigma_0$ and a time and wind direction dependent PM10 concentration $\delta_{est}$ of the air brought above the city by the wind, it is reasonably expected to perform even better.

The model demonstrates how daily PM10 pollution concentration values depend on the weather condition of the day and of those of the previous days. Thus, because meteorological parameters can be forecast one or two days ahead with a high accuracy, the proposed model is a valid and inexpensive tool for predicting a few days in advance the possibility of air pollution crisis. The model could be adopted by city planners and managers to establish regulations able to control the pollutant factors of Naples.

Likely, the proposed semi-empirical model, due to its basic general equations, can be easily adapted to simulate air pollution concentration values in many other localities.

Zhang, Y., Pun, B., Wu, S.Y., Vijayaraghavan, K., Seigneur, C., 2004. Application and evaluation of two air quality models for particulate matter for a southeastern U.S. episode. Journal of the Air & Waste Management Association,54,1478-1493.

## Web sites

- www.campanialive.it
- www.arpacampania.it
- http://www.guardian.co.uk/environment/gallery/2012/dec/05/60-years-great-smog-london-in-pictures#/?picture=400439580&index=6
- http://library.thinkquest.org/C003603/english/fogandmist/casestudies.shtml#2
- www.wetterzentrale.de
- www.westwind.ch

| Inversion Index (°C) | Air pollution severity |
|---|---|
| $\Delta\theta < -2.5$ | Very low |
| $-2.5 \leq \Delta\theta < 0.5$ | Low |
| $0.5 \leq \Delta\theta < 2.5$ | Moderate |
| $2.5 \leq \Delta\theta < 5.0$ | High |
| $\Delta\theta \geq 5.0$ | Exceptional |

Table 1: Qualitative relation between the air temperature inversion index between the meteorological stations of Sant'Elmo and San Marcellino and the air pollution severity in Naples.

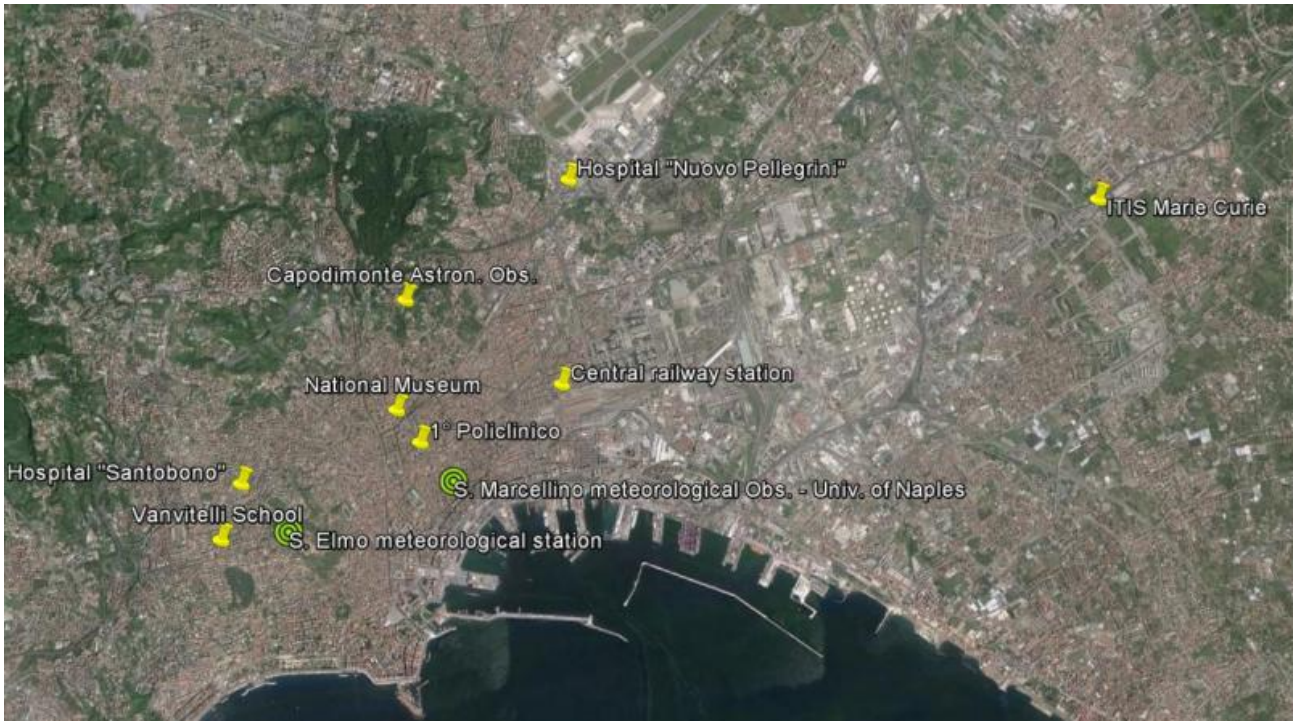

Figure 1: Map of the city of Naples with the position of the locations of PM10 stations and of the meteorological observatories in San Marcellino e Sant'Elmo.

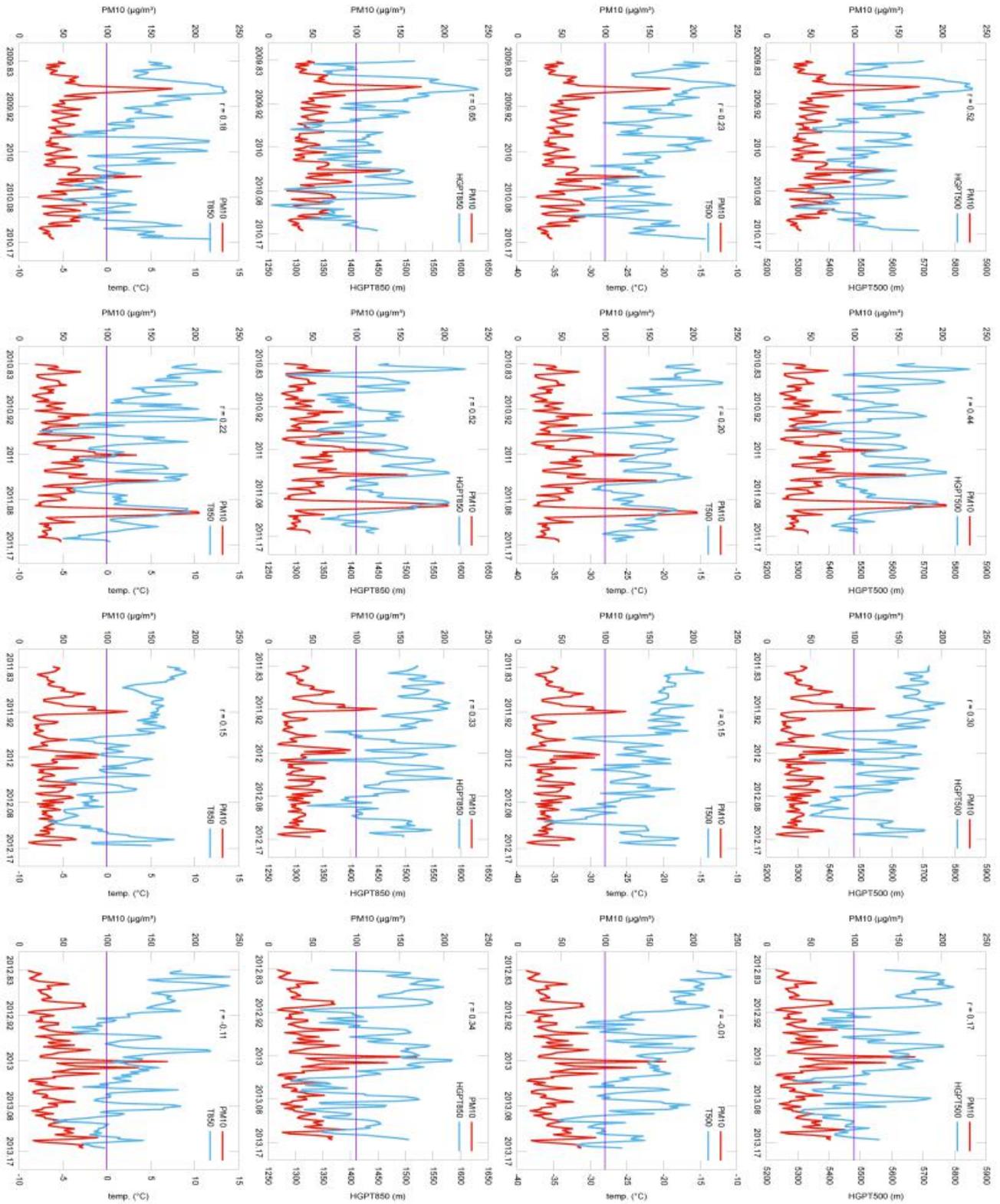

Figure 2: Daily PM10 concentration levels (red) measured in Naples from November to February for the years from 2009 to 2013 against the geopotential heights and temperature at 850 hPa and at 500 hPa above the city of Naples. Each panel also reports the cross correlation coefficient r between the two records.

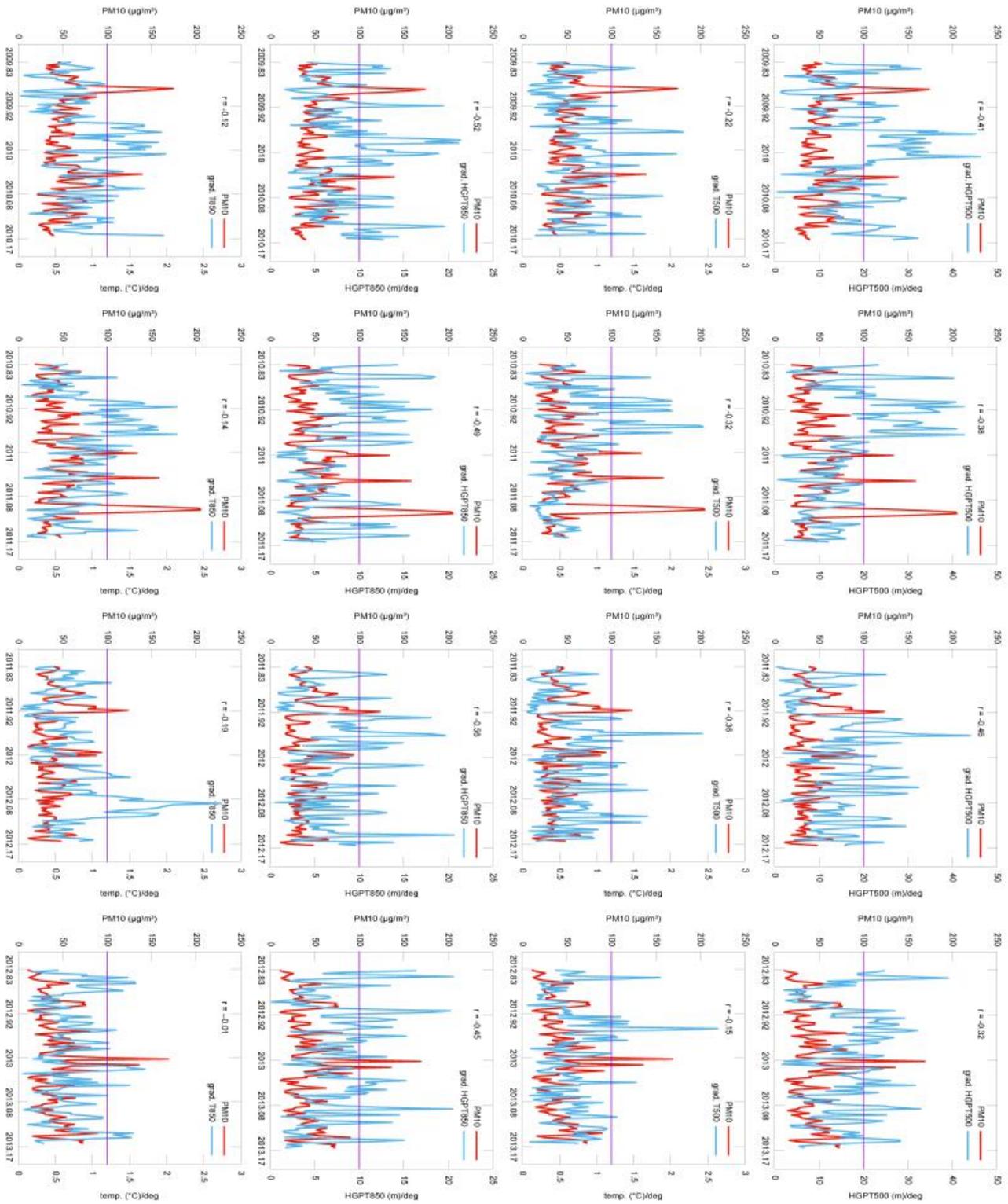

Figure 3: Daily PM10 concentration levels (red) measured in Naples from November to February for the years from 2009 to 2013 against the gradients of geopotential heights and temperature at 850 hPa and at 500 hPa above the city of Naples. Each panel also reports the cross correlation coefficient r between the two records.

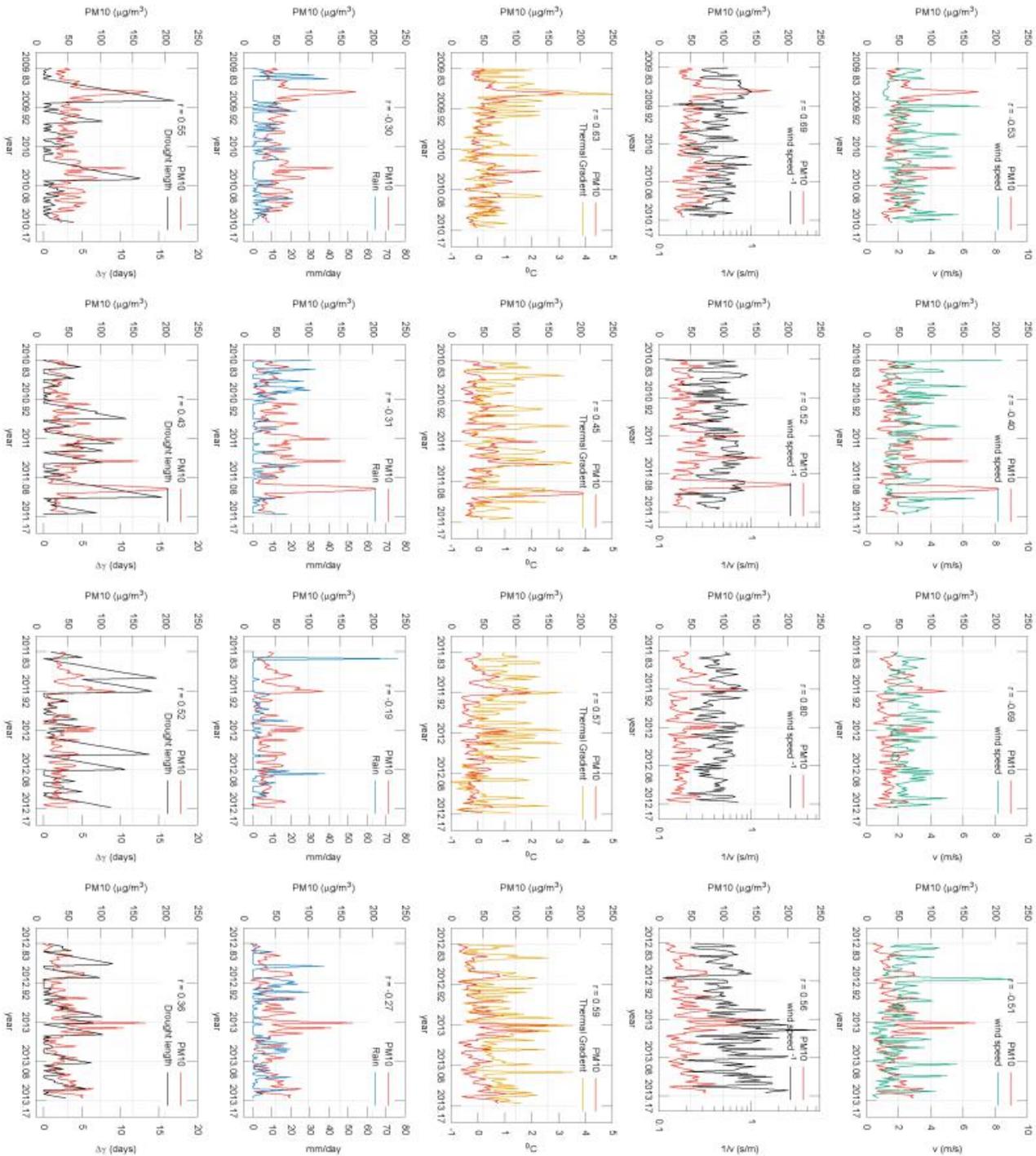

Figure 4: Daily PM10 concentration levels (red) measured in Naples from November to February for the years from 2009 to 2013 against the various meteorological parameters measured at the observatory of San Marcellino in the center on Naples (wind speed and its reciprocal, daily rain, number of past draught days) and the thermal gradient calculated using the minimum daily temperature measured at both the observatory of San Marcellino and Sant'Elmo. Each panel also reports the cross correlation coefficient r between the two records.

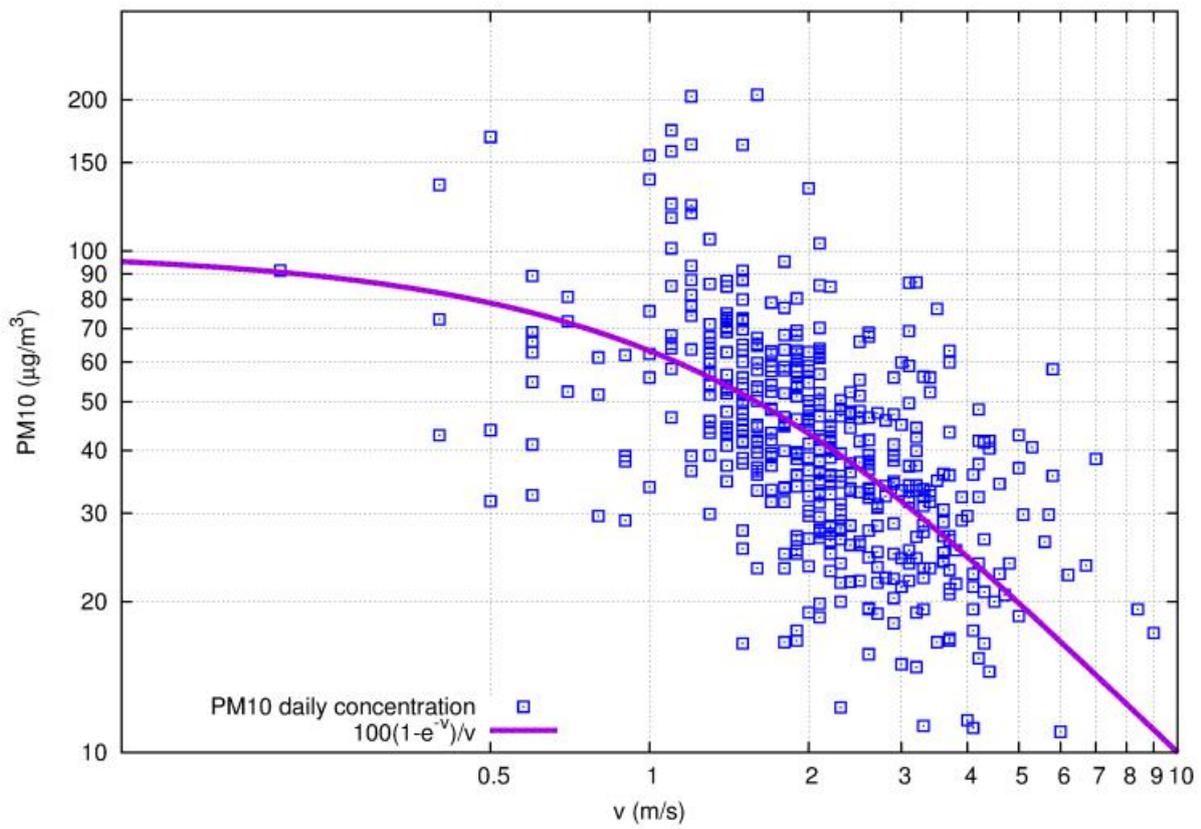

Figure 5: Daily PM10 concentration levels measured in Naples from November to February for the years from 2009 to 2013 against the wind speed. The data are fit with Eq. 8.

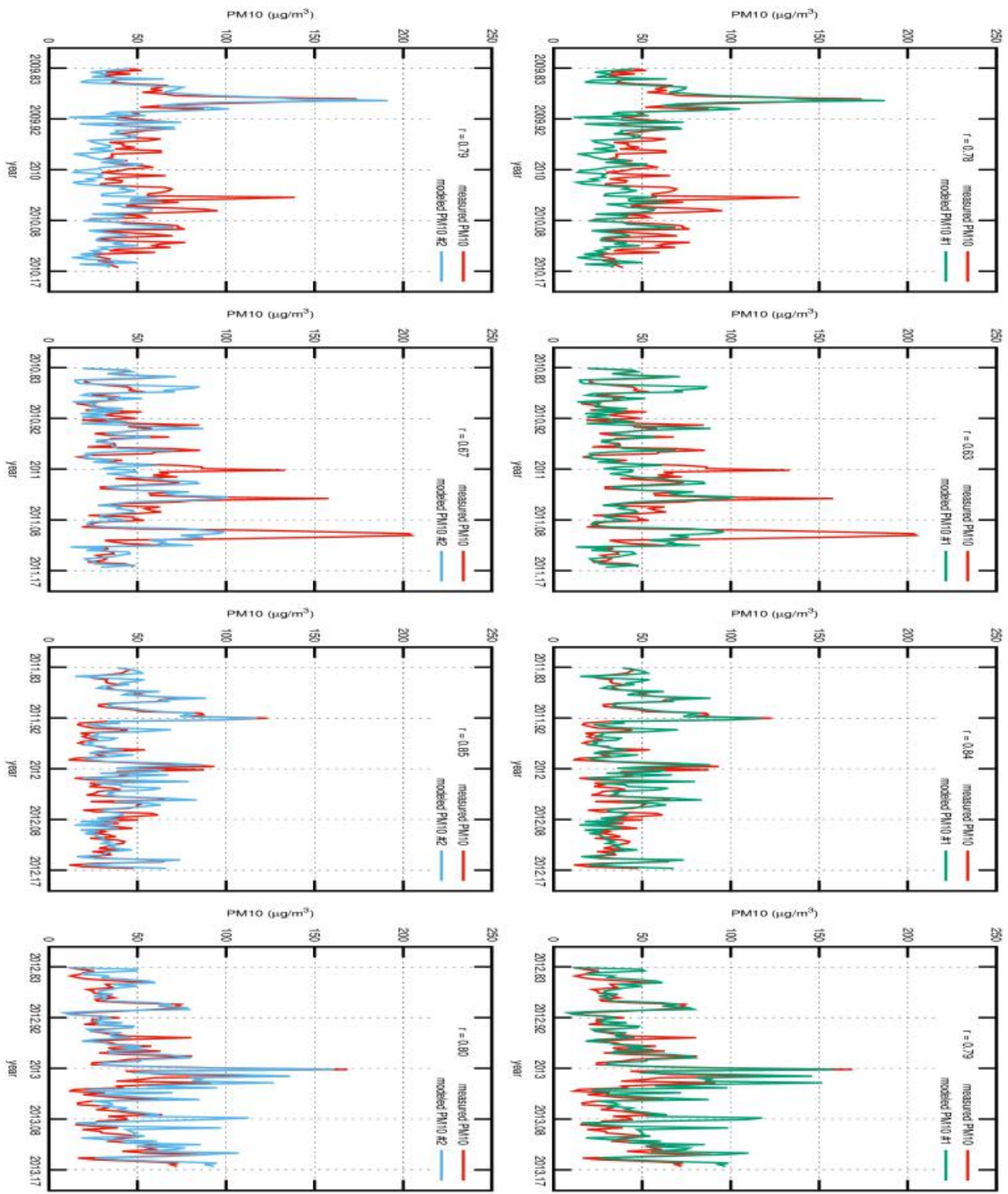

Figure 6: Daily PM10 concentration levels (red) measured in Naples from November to February for the years from 2009 to 2013 against the prediction of the PM10 model described by the Eqs. 1-7. The upper panel model #1 uses for $\delta(0)$ the $\delta_{eq}$ value calculated by the model for the previous day. The lower panel model #2 uses for $\delta(0)$ the PM10 value measured the previous day. Each panel also reports the cross correlation coefficient r between the two records.